\documentclass[letterpaper,12pt,amsfonts,amssymb]{article}

\usepackage{graphics,graphicx,psfrag,color,float}
\usepackage{epsfig}
\usepackage{amsmath}
\usepackage{amssymb}

\begin{document}
\newcommand{\hrho}{\widehat{\rho}}
\newcommand{\hsigma}{\widehat{\sigma}}
\newcommand{\homega}{\widehat{\omega}}
\newcommand{\hI}{\widehat{I}}
\newcommand*{\spr}[2]{\langle #1 | #2 \rangle}
\newcommand*{\bbN}{\mathbb{N}}
\newcommand*{\bbR}{\mathbb{R}}
\newcommand*{\cA}{\mathcal{A}}
\newcommand*{\cB}{\mathcal{B}}
\newcommand*{\barpi}{\overline{\pi}}
\newcommand*{\barP}{\overline{P}}
\newcommand*{\eps}{\varepsilon}
\newcommand*{\id}{I}
\newcommand{\orho}{\overline{\rho}}
\newcommand{\omu}{\overline{\mu}}
\newcommand*{\half}{{\frac{1}{2}}}
\newcommand*{\ket}[1]{| #1 \rangle}
\newcommand{\trho}{{\widetilde{\rho}}_n^\gamma}
\newcommand{\ttrho}{{\widetilde{\rho}}}
\newcommand{\tsigma}{{\widetilde{\sigma}}}
\newcommand{\tpi}{{\widetilde{\pi}}}
\newcommand*{\bra}[1]{\langle #1 |}
\newcommand*{\proj}[1]{\ket{#1}\bra{#1}}
\newcommand{\otrho}{{\widetilde{\rho}}_n^{\gamma 0}}
\newcommand{\be}{\begin{equation}}
\newcommand{\bea}{\begin{eqnarray}}
\newcommand{\eea}{\end{eqnarray}}
\newcommand{\tr}{\mathrm{Tr}}
\newcommand*{\Hmin}{H_{\min}}
\newcommand{\rank}{\mathrm{rank}}
\newcommand{\tends}{\rightarrow}
\newcommand{\uS}{\underline{S}}
\newcommand{\oS}{\overline{S}}
\newcommand{\uD}{\underline{D}}
\newcommand{\oD}{\overline{D}}
\newcommand{\ee}{\end{equation}}
\newcommand{\supp}{\rm{supp}}
\newcommand{\n}{{(n)}}
\newtheorem{definition}{Definition}
\newtheorem{theorem}{Theorem}
\newtheorem{proposition}{Proposition}
\newtheorem{lemma}{Lemma}
\newtheorem{defn}{Definition}
\newtheorem{corollary}{Corollary}
\newcommand{\qed}{\hspace*{\fill}\rule{2.5mm}{2.5mm}}
\newcommand{\beq}{\begin{equation}}
\newcommand{\enq}{\end{equation}}
\newcommand{\beqa}{\begin{eqnarray}}
\newcommand{\enqa}{\end{eqnarray}}
\newcommand{\beqan}{\begin{eqnarray*}}
\newcommand{\enqan}{\end{eqnarray*}}
\newcommand{\mycal}[1]{\mathcal #1}
\newcommand\figcaption{\def\@captype{figure}\caption}
\newcommand{\mbm}[1]{\pmb{#1}}
\newcommand{\transpose}{\mathsf{T}}
\newcommand{\nvec}[1]{\mathrm{vec}\left(#1\right)}
\newcommand{\nvecsq}[1]{\mathrm{vec}\left[#1\right]}
\newcommand{\mexp}[1]{\mathsf{E}\left\{ #1 \right\}}
\newcommand{\frel}{quantum $f$-relative entropy~}
\newcommand{\frelns}{quantum $f$-relative entropy}
\newcommand{\fent}{quantum $f$-entropy~}
\newcommand{\fentns}{quantum $f$-entropy}
\newcommand*{\braket}[2]{\left\langle #1 \right| \left. #2 \right\rangle}
\newcommand*{\brakett}[3]{\left\langle #1 \right| \left. #2 \right. \left| #3 \right\rangle}
\newcommand{\define}{\stackrel{\Delta}{=}}
\newcommand{\bit}{\begin{itemize}}
\newcommand{\eit}{\end{itemize}}

\newenvironment{proof}{\noindent{\it Proof}\hspace*{1ex}}{\qed\medskip}
\def\reff#1{(\ref{#1})}

\title{On the quantum $f$-relative entropy and generalized data processing inequalities}

\author{{\bf Naresh Sharma} \\
Tata Institute of Fundamental Research \\
Mumbai 400 005, India \\
Email: \texttt{nsharma@tifr.res.in}
}

\date{\today}

\maketitle

\begin{abstract}
We study the fundamental properties of the \frelns, where $f(\cdot)$ is an operator
convex function. We give the equality conditions under various properties including
monotonicity and joint convexity, and these conditions are more general than,
since they hold for a class of operator convex functions, and different for
$f(t) = -\ln(t)$ from, the previously known conditions.
The \fent is defined in terms of the \frel and we study its properties giving 
the equality conditions in some cases. We then show that the $f$-generalizations of
the Holevo information, the entanglement-assisted capacity, and the coherent information
also satisfy the data processing inequality, and give the equality conditions for
the $f$-coherent information.
\end{abstract}

\section{Introduction}

Quantum entropy is central to the study of information processing in
quantum mechanical systems (see \cite{petz-book,nielsen-chuang}
and references therein). The von Neumann entropy for a density matrix $\rho$,
a positive semi-definite matrix ($\rho \geq 0$) with unit trace
($\tr(\rho)=1$), is given
by
\beq
S(\rho) = -\tr\left[\rho \ln(\rho) \right].
\enq
Schumacher's quantum noiseless channel coding theorem gives an information-theoretic
interpretation of this quantity \cite{schu-noiseless-1995}. Lieb and Ruskai showed that
the von Neumann entropy satisfies, among other inequalities, the strong sub-additivity
\cite{lieb-ruskai-ssa0-1973,lieb-ruskai-ssa-1973} given by
\beq
\label{ssa}
S(\rho_{ABC}) + S(\rho_B) \leq S(\rho_{AB}) + S(\rho_{BC}),
\enq
where $ABC$ is the composite system consisting of subsystems $A$, $B$, and $C$
with the density matrix $\rho_{ABC}$, and the density
matrices of subsystem(s) is obtained by tracing out other subsystem(s). For example,
$\rho_{AB} = \tr_C(\rho_{ABC})$. The equality conditions for the strong sub-additivity
were given by Hayden \emph{et al} \cite{equality-ssa-2004}.

Umegaki defined the quantum relative entropy of $\rho$ to
$\sigma$ as \cite{umegaki-1962}
\beq
S(\rho || \sigma) = \tr\left\{ \rho \left[ \ln(\rho) - \ln(\sigma) \right] \right\},
\enq
where $\rho$, $\sigma$ are density matrices.
Lindblad proved the monotonicity of the quantum relative entropy, which is stated as
\beq
S(\rho_{AB} || \sigma_{AB}) \geq S(\rho_A || \sigma_A),
\enq
where $\rho_A = \tr_B(\rho_{AB})$ and $\sigma_A = \tr_B(\sigma_{AB})$
\cite{lindblad1975}.
The equality conditions for the quantum relative entropy under monotonicity
were given by Petz \cite{petz-equality-rentr-1986} and Ruskai
\cite{ruskai-equality-2002}, and these conditions are equivalent though not the same
and are obtained using different approaches. Ibinson, Linden, and
Winter later showed that monotonicity under restrictions is 
the only general inequality satisfied by quantum relative entropy \cite{winter-mono-2005}.
The joint convexity property of the quantum relative entropy is stated
for $0 \leq \lambda \leq 1$, and density matrices $\sigma_i$, $\rho_i$, $i=1,2$, as
\beq
S\left[ \lambda \rho_1 + (1 - \lambda) \rho_2 ||
\lambda \sigma_1 + (1 - \lambda) \sigma_2 \right] \leq
\lambda S(\rho_1 || \sigma_1) + (1 - \lambda) S(\rho_2 || \sigma_2).
\enq
Ruskai \cite{ruskai-equality-2002} gave the equality conditions under joint convexity.
Ruskai describes an elegant way of deducing strong sub-additivity and
joint convexity from monotonicity in Ref. \cite{ruskai-equality-2002}.

\subsection{Operator convex functions}

If $A$ is Hermitian and has a spectral decomposition given by
\beq
A = \sum_{i} \alpha_i \ket{i} \bra{i},
\enq
then the matrix valued function $f(A)$ is defined as
\beq
f(A) = \sum_i f(\alpha_i) \ket{i} \bra{i},
\enq
where we have implicitly assumed that the spectrum of $A$ lies in the domain of $f$.

A real valued function $f(\cdot)$ is said to be operator convex if for all Hermitian
matrices $A$ and $B$, and $0 \leq \lambda \leq 1$,
\beq
f \left[\lambda A + (1-\lambda) B \right] \leq \lambda f(A) + (1-\lambda) f(B).
\enq
It is easy to see that if $f(t)$ is operator convex, then so is $g(t) = f(t-t_0)$
for some $t_0 \in {\mathbb{R}}$ assuming that $t-t_0$ is in the domain of $f(\cdot)$.

It follows from Theorem V.4.6 in Ref. \cite{bhatia-1997} that for a non-affine operator convex
function $f(\cdot)$ on $(-a+t_0,a+t_0)$, there exists a unique probability density
function $p_X(x)$ defined on $x \in [-a,a]$ such that for $d > 0$,
\beq
\label{op_cx}
f(t) = b + c t + d\int_{-a}^a \frac{(t-t_0)^2}{a^2-(t-t_0)x} p_X(x) dx.
\enq
Choosing $t_0=a$, $b=1-\ln(a)$, $c=-1/a$, $d=1/2$, and
\beq
p_X(x) = \left\{
\begin{array}{ll}
- \frac{2}{a^2} x, & -a \leq x \leq 0, \\
0, &  \mbox{otherwise},
\end{array}
\right.
\enq
we get
\beq
f(t) = -\ln(t), ~~ t \in (0,2a).
\enq

\ \\
\noindent {\bf Definition}: An operator convex function is said to be \emph{diffused} if the
probability density function, $p_X(x)$, in Eq. (\ref{op_cx}) is strictly positive a.e.
in a subinterval of $[-a,a]$ that has $x=0$ as the interior or the boundary point. \\

The function $f(t) = -\ln(t)$ clearly belongs to this class of operator convex functions
since $p_X(x)$ is strictly positive $\forall ~ x \in [-a,0)$ and $x=0$ is a boundary point
of this interval.

We mention the operator Jensen's inequality that we shall use more than
once in this paper.

\begin{theorem}
\label{jensen_op}
(Hansen and Pedersen \cite{hansen-1982}) Let $E_i$, $i=1,...,n$, be a set of
matrices satisfying
\beq
\sum_{i=1}^n E_i^\dagger E_i = {\mathrm{I}}.
\enq
Then, for Hermitian matrices $\phi_i$, $i = 1,...,n$, with bounded
spectra, and an operator convex function $f(\cdot)$,
\beq
\label{jensen}
f \left( \sum_{i=1}^n E_i^\dagger \phi_i E_i \right) \leq
\sum_{i=1}^n E_i^\dagger f(\phi_i) E_i.
\enq
\end{theorem}

\subsection{Quantum $f$-relative entropy}

Let $A$ be a $m \times n$ matrix
\beq
A = [a_{ij}], ~~ i = 1,...,m, ~~~ j = 1,...n,
\enq
where $a_{ij}$ is the $(i,j)$th entry of $A$.
We denote conjugate, transpose, and conjugate transpose of $A$ by
$A^*$, $A^\transpose$, and $A^\dagger$ respectively. One can
associate a vector with matrix $A$, denoted by $\nvec{A}$, whose
$[n(i-1)+j]$th entry, denoted by $\nvec{A}_{n(i-1)+j}$ is given by
\beq
\nvec{A}_{n(i-1)+j} = a_{ij}.
\enq
One can, of course, construct $A$ back from $\nvec{A}$.
An identity that we shall frequently employ is \cite{watrous-notes}
\beq
\label{duminy0}
\nvec{ABC} = (A \otimes C^\transpose) \nvec{B},
\enq
where $A$, $B$, $C$ are matrices with appropriate dimensions.

It is well known that many properties of the quantum relative entropy are not central
to the $\ln(\cdot)$ used in its definition and a more general definition of quantum
relative entropy is studied in \cite{petz-book,anna-equality-2009}. In the classical
case, the $f$-generalization of the classical relative entropy was
studied by Csisz\'{a}r \cite{csiszar-f-1970}.

The \frel for strictly positive $\rho$ and
$\sigma$ is defined as
\beq
\label{frentropy1}
S_f(\rho || \sigma) = \nvec{\sqrt{\rho}}^\dagger
f \left[ \sigma \otimes \left(\rho^{-1}\right)^\transpose \right] \nvec{\sqrt{\rho}},
\enq
where $f(\cdot)$
is an operator convex function. We shall implicitly assume that the domain of $f(\cdot)$
is contained in $(0,a)$ for some finite $a > 0$.
Note that we don't impose the condition that $\rho$ and $\sigma$ have unit trace.
Let the
spectral decompositions of $\rho$ and $\sigma$ in Eq. (\ref{frentropy1}) be given by
\beqa
\label{sdrho}
\rho & = & \sum_{i=1}^d p_i \ket{i_\rho} \bra{i_\rho}, \\
\label{sdsigma}
\sigma & = & \sum_{j=1}^d q_j \ket{j_\sigma} \bra{j_\sigma},
\enqa
where $d$ is the dimension of the Hilbert space that describes $\rho$ and $\sigma$.
Using Eqs. (\ref{duminy0}), (\ref{frentropy1}), (\ref{sdrho}), and (\ref{sdsigma}),
we can also write the \frel as
\beqa
\label{frentropy5}
S_f(\rho || \sigma)  & = & \nvec{{\mathrm{I}}}^\dagger \left[{\mathrm{I}} \otimes
\left(\sqrt{\rho}\right)^*\right] f \left[ \sigma \otimes \left(\rho^{-1}\right)^\transpose
\right]
\left[{\mathrm{I}} \otimes \left(\sqrt{\rho}\right)^\transpose\right] \nvec{{\mathrm{I}}}, \\
\label{frentropy4}
& = & \sum_{i,j=1}^d p_i f\left(\frac{q_j}{p_i} \right)
|\braket{i_\rho}{j_\sigma}|^2, \\
\label{frentropy2}
& = & \sum_{j=1}^d \bra{j_\sigma} \sqrt{\rho}
f(q_j \rho^{-1}) \sqrt{\rho} \ket{j_\sigma}, \\
\label{frentropy3}
& = & \sum_{i=1}^d p_i \brakett{i_\rho}
{f\left(\frac{\sigma}{p_i}\right)}{i_\rho},
\enqa
where ${\mathrm{I}}$ is the Identity matrix whose dimensions, if unspecified, would be
apparent from the context.

\subsection{Overview}

We follow the ``vec" notation throughout this paper as was used in our definition of the \frel
in Eq. (\ref{frentropy1}) as opposed to the linear super-operators in \cite{petz-book,
petz-quasi-entr-1986,anna-equality-2009}. We shall see that this notation along with the
operator Jensen's inequality in Theorem \ref{jensen_op} gives alternate and more
accessible proofs of many inequalities and equality conditions.

We note here that the \frel defined in Eq. (\ref{frentropy1}) is a special case of quantum quasi
relative entropy defined by Petz \cite{petz-book,petz-quasi-entr-1986}. 
However, we shall see that we arrive at the equality conditions
under certain properties for a class of
operator convex functions, which are more general than, and different in case of $f(t) = -\ln(t)$
from, those given by Petz \cite{petz-equality-rentr-1986} and Ruskai
\cite{ruskai-equality-2002}.

We note here that since we define the \frel for strictly positive matrices, in some cases
in this paper, it shall put implicit restrictions. For example, when we deal with
the \frel after processing, i.e., $S_f[{\mathcal{E}}(\rho) || {\mathcal{E}}(\sigma)]$,
where ${\mathcal{E}}(\cdot)$ is a quantum operation,
we shall implicitly assume that ${\mathcal{E}}(\rho)$ and ${\mathcal{E}}(\sigma)$
are strictly positive, which puts restrictions on the choices of $\rho$, $\sigma$, and
${\mathcal{E}}(\cdot)$. A way out could have been to extend the definition of the \frel for
positive semi-definite matrices. This could be accomplished by defining the
terms of the form $f(0)$, $0 \times f(0/0)$, and $0 \times f(a/0)$, $a > 0$.
But we refrain from doing that in this paper since we deal with a class of operator
convex functions and leave the extension of the definition to the time
when a specific choice of the function $f(\cdot)$ is made in Eq. (\ref{frentropy1}),
which we won't do in this paper.

We define the \fent in terms of the \frel and study some of its properties and give
the equality conditions for some cases. We also show the $f$-generalizations of
some well-known quantum information-theoretic quantities also satisfy the
data processing inequalities as is the case for $f(t) = -\ln(t)$.

\section{Properties of the \frelns}

We now list some useful properties of the \frelns.
\begin{lemma}
For strictly positive $\rho$ and $\sigma$, the following properties hold:
\bit
\item[(i)] The \frel is invariant under Unitary transformation, i.e.,
\beq
S_f\left(U \rho U^\dagger || U \sigma U^\dagger\right) = S_f(\rho || \sigma),
\enq
where $U^\dagger U = {\mathrm{I}}$.
\item[(ii)] For any strictly positive $\kappa$,
\beq
S_f(\rho \otimes \kappa || \sigma \otimes \kappa) = S_f(\rho || \sigma).
\enq
\item[(iii)] For any scalar $c > 0$,
\beq
\frac{1}{c} S_f( c\rho || c \sigma) = S_f(\rho || \sigma).
\enq
\eit
\end{lemma}
\begin{proof}
These properties follow easily from Eq. (\ref{frentropy4}) and we omit the proof.
\end{proof}

\subsection{Monotonicity}

Petz \cite{petz-quasi-entr-1986}, Nielsen and Petz \cite{nielsen-2005} provide an elegant
proof the monotonicity of the \frelns. We restate their proof in the ``vec" notation.
\begin{lemma} (Petz \cite{petz-quasi-entr-1986}, Nielsen and Petz \cite{nielsen-2005})
Let $\rho_{AB}$ and $\sigma_{AB}$
be two strictly positive matrices in the composite system consisting of systems $A$ and $B$,
and let $\rho_A = \tr_B(\rho_{AB})$ and $\sigma_A = \tr_B(\sigma_{AB})$. Then
\beq
\label{mono}
S_f(\rho_{AB} || \sigma_{AB}) \geq S_f(\rho_A || \sigma_A).
\enq
\end{lemma}
\begin{proof} 
Let us assume that there exists a matrix $V$ such that
\beqa
\label{dummy1}
V \nvec{\sqrt{\rho_A}} & = & \nvec{\sqrt{\rho_{AB}}}, \\
\label{dummy2}
V^\dagger V & = & {\mathrm{I}}, \\
\label{dummy3}
V^\dagger \left[ \sigma_{AB} \otimes \left( \rho_{AB}^{-1} \right)^\transpose \right] V
& = & \sigma_A \otimes \left( \rho_A^{-1} \right)^\transpose.
\enqa
To show that such a $V$ does exist, let us consider a linear super-operator
${\mathcal U}(\cdot)$ such that
\beq
{\mathcal U}(X) = \left( X \rho_A^{-1/2} \otimes {\mathrm{I}} \right) \sqrt{\rho_{AB}}.
\enq
Its adjoint is given by
\beq
{\mathcal U}^\dagger(Y) = \tr_B \left[ Y \sqrt{\rho_{AB}} \left(\rho_A^{-1/2}
\otimes {\mathrm{I}} \right) \right].
\enq
That this is indeed the adjoint is evident from
\beq
\langle {\mathcal U}^\dagger(Y),X \rangle = \langle Y,{\mathcal U}(X) \rangle,
\enq
where $\langle E,F \rangle = \tr(E^\dagger F)$ is the Hilbert-Schmidt inner product.
Let us associate a matrix $V$ with ${\mathcal U}$ such that
\beqa
\nvecsq{{\mathcal U}(X)} & = & V \nvec{X}
\enqa
and hence,
\beqa
\nvecsq{{\mathcal U}^\dagger(Y)} & = & V^\dagger \nvec{Y}.
\enqa
Note that since
\beqa
{\mathcal U}\left( \sqrt{\rho_A} \right) & = & \sqrt{\rho_{AB}}, \\
{\mathcal U}^\dagger\left[{\mathcal U}(X)\right] & = & X, \\
{\mathcal U}^\dagger\left[\sigma_{AB} {\mathcal U}(X) \rho_{AB}^{-1} \right]
& = & \sigma_A X \rho_A^{-1},
\enqa
Eqs. (\ref{dummy1}), (\ref{dummy2}) and (\ref{dummy3}) must hold. We now have
\beqa
S_f(\rho_{AB} || \sigma_{AB}) & = & \nvec{\sqrt{\rho_{AB}}}^\dagger
f \left[ \sigma_{AB} \otimes \left( \rho_{AB}^{-1} \right)^\transpose \right] 
\nvec{\sqrt{\rho_{AB}}} \\
\label{dummy4}
& = & \nvec{\sqrt{\rho_{A}}}^\dagger V^\dagger
f \left[ \sigma_{AB} \otimes \left( \rho_{AB}^{-1} \right)^\transpose \right]
V \nvec{\sqrt{\rho_{A}}} \\
\label{app_jensen}
& \geq & \nvec{\sqrt{\rho_{A}}}^\dagger f \left\{ V^\dagger
\left[ \sigma_{AB} \otimes \left( \rho_{AB}^{-1} \right)^\transpose \right] V \right\}
\nvec{\sqrt{\rho_{A}}} \\
& = & \nvec{\sqrt{\rho_{A}}}^\dagger f
\left[ \sigma_{A} \otimes \left( \rho_{A}^{-1} \right)^\transpose \right]
\nvec{\sqrt{\rho_{A}}} \\
& = & S_f(\rho_A || \sigma_A),
\enqa
where Eq. (\ref{app_jensen}) follows from Eq. (\ref{dummy4}) by using the operator
Jensen's inequality in Eq. (\ref{jensen}) with $n=1$ and $E_1 = V$.
\end{proof}

We now give the conditions for the equality in Eq. (\ref{mono})
for a class of operator convex functions.

\begin{lemma}
\label{lemma-equality}
For a non-affine and diffused operator convex function $f(\cdot)$,
positive $\rho_{AB}$, $\sigma_{AB}$, $\rho_A = \tr_B(\rho_{AB})$ and
$\sigma_A =$ $\tr_B(\sigma_{AB})$, the equality in
\beq
\label{equality0}
S_f \left( \rho_{AB} || \sigma_{AB} \right) \geq
S_f \left(\rho_A || \sigma_A \right)
\enq
holds if and only if
\beq
\label{f-eq}
\tr\left(\sigma_{AB}^{\imath t} \rho_{AB}^{-\imath t+1} \right) =
\tr\left(\sigma_A^{\imath t} \rho_A^{-\imath t+1} \right), ~~ \forall ~ t \in \mathbb{C},
\enq
where $\imath = \sqrt{-1}$.
\end{lemma}
\begin{proof}
Let
\beq
\gamma_{AB} = \sigma_{AB} \otimes \left(\rho_{AB}^{-1}\right)^\transpose,
\enq
\beq
\gamma_A = \sigma_A \otimes \left(\rho_A^{-1}\right)^\transpose.
\enq
Since
\beq
S_f \left( \rho_{AB} || \sigma_{AB} \right) =
\nvec{\sqrt{\rho_{AB}}}^\dagger f(\gamma_{AB}) \nvec{\sqrt{\rho_{AB}}},
\enq
\beq
S_f (\rho_A || \sigma_A) =
\nvec{\sqrt{\rho_{A}}}^\dagger f(\gamma_A) \nvec{\sqrt{\rho_{A}}},
\enq
hence, using Eq. (\ref{op_cx}), we get
\beqa
S_f \left( \rho_{AB} || \sigma_{AB} \right) - S_f (\rho_A || \sigma_A)
& = & d \int_{(-a,a)} r(x) p_X(x) dx,
\enqa
where $d > 0$ since $f(\cdot)$ is non-affine, and
\beqa
r(x) & = & \nvec{\sqrt{\rho_{AB}}}^\dagger \left(\gamma_{AB} - t_0
{\mathrm{I}}\right)^2
\left[ a^2 {\mathrm{I}} - (\gamma_{AB} - t_0 {\mathrm{I}}) x \right]^{-1}
\nvec{\sqrt{\rho_{AB}}} - \\
& & ~~~~ \nvec{\sqrt{\rho_A}}^\dagger \left(\gamma_A - t_0 {\mathrm{I}}\right)^2
\left[ a^2 {\mathrm{I}} - \left(\gamma_A - t_0 {\mathrm{I}}\right) x \right]^{-1}
\nvec{\sqrt{\rho_A}}.
\enqa
Since $f(\cdot)$ is diffused, let us assume that
$p_X(x)$ in Eq. (\ref{op_cx}) is strictly positive a.e. over $\acute{K}$, a subinterval of $[-a,a]$
with the point $x=0$ as the interior or the boundary point.
Consider $K$, a subinterval of $\acute{K}$,
with the point $x=0$ as the interior or the boundary point, and that does not contain the
points at which $a^2 {\mathrm{I}} - (\gamma_{AB} - t_0 {\mathrm{I}}) x$
or $a^2 {\mathrm{I}} - (\gamma_A - t_0 {\mathrm{I}}) x$ become singular.
Such a $K$ is always possible since there are finite values of $x$ for which
the above two matrices become singular and these matrices are non-singular in the
neighborhood of $x=0$.

Now note that the function $g(t) = (t-t_0)^2/\left[ a^2 - (t-t_0)x \right]$ is operator
convex. To see this, first note that $1/t$, $t \in (0,\infty)$ is operator convex
(see Corollary V.2.6 in Ref. \cite{bhatia-1997} and Ref. \cite{nielsen-2005}).
It follows that $h(t) = 1/(a^2t-x)$ is also operator convex in $t \in (x/a^2,\infty)$.
Using Lemma \ref{lemma-g} in Sub-section \ref{sub-convexity}, it follows that
$t ~ h(1/t) = t^2/(a^2-tx)$,
$t \in (0,a^2/x)$, is also operator convex, which implies that $g(t)$ is also operator convex
in the neighborhood of $x=0$. Hence, using monotonicity, $r(x) \geq 0$.
For the equality to hold,
\beqa
0 = \int_{(-a,a)} r(x) p_X(x) dx \geq \int_{K} r(x) p_X(x) dx \geq 0.
\enqa
Hence, $r(x) p_X(x) = 0$ a.e. over $K$. Since $p_X(x) > 0$ a.e. over $K$, hence,
$r(x) = 0$ a.e. over $K$. But since $r(x)$ is continuous in $K$, hence,
$r(x) = 0$ over $K$.
Hence, the coefficients of the Taylor series expansion of $r(x)$ around $x=0$
must be zero. (If $x=0$ is the boundary point of $K$, then we shall take the Taylor
series expansion of $r(x)$ around $x=\epsilon \in K$, where $|\epsilon|$ is arbitrarily small
and then in the limit of $\epsilon$ approaching zero, we shall arrive at the same
conclusions as below.) Equating the Taylor series coefficients to be zero (and
taking appropriate limits if $x=0$ is the boundary point of $K$), we have
\beq
\nvec{\sqrt{\rho_{AB}}}^\dagger \left(\gamma_{AB} - t_0 {\mathrm{I}} \right)^n 
\nvec{\sqrt{\rho_{AB}}} =
\nvec{\sqrt{\rho_A}}^\dagger \left(\gamma_A - t_0 {\mathrm{I}} \right)^n 
\nvec{\sqrt{\rho_A}}, ~~ \forall ~ n \geq 2.
\enq
The above equation is trivially true for $n=0,1$, and it follows that the above equation
is equivalent to
\beqa
\nvec{\sqrt{\rho_{AB}}}^\dagger \gamma_{AB}^n  \nvec{\sqrt{\rho_{AB}}} & = &
\nvec{\sqrt{\rho_A}}^\dagger \gamma_A^n  \nvec{\sqrt{\rho_A}}, ~~ \forall ~ n \geq 0,
~~~~ \\
\label{dummy51}
\tr\left( \sigma_{AB}^n \rho_{AB}^{-n+1} \right)
& = & \tr\left(\sigma_A^n \rho_{A}^{-n+1} \right),
~~ \forall ~ n \geq 0.
\enqa
Let the spectral decompositions of the matrices in the above equation be given by
$\rho_{AB} = $ $\sum_k p_k \ket{k^\rho_{AB}} \bra{k^\rho_{AB}}$,
$\sigma_{AB} = \sum_k q_k \ket{k^\sigma_{AB}} \bra{k^\sigma_{AB}}$,
$\rho_A = \sum_k r_k \ket{k^\rho_A} \bra{k^\rho_A}$, and
$\sigma_A = \sum_k s_k \ket{k^\sigma_A} \bra{k^\sigma_A}$. Substituting in the
above equation, we get
\beq
\sum_{k,j} \left( \frac{q_j}{p_k} \right)^n p_k \left|
\braket{j^\sigma_{AB}}{k^\rho_{AB}} \right|^2 =
\sum_{k,j} \left( \frac{s_j}{r_k} \right)^n r_k \left|
\braket{j^\sigma_A}{k^\rho_A} \right|^2, ~~ \forall ~ n \geq 0.
\enq
Now consider the terms in the LHS such that $q_j/p_k$ $=\max_{k,j} q_j/p_k$
and in the RHS such that $s_j/r_k$ $=\max_{k,j} s_j/r_k$.
It is clear that for large $n$, these two set of terms dominate all other terms,
and since LHS $=$ RHS,
$\max_{k,j} q_j/p_k = \max_{k,j} s_j/r_k$, and the sum of their coefficients in
the LHS must be the same as their sum in the RHS. Subtracting these two sets from
both sides and arguing similarly for the maximum in the pruned summations
and continuing till no term is left, it follows that
Eq. (\ref{dummy51}) amounts to Eq. (\ref{f-eq}).
\end{proof}

We note that the conditions for the equality in Eq. (\ref{mono}) for $f(t) = -\ln(t)$ were given by
Petz \cite{petz-equality-rentr-1986} as
\beq
\label{petz-eq}
\sigma_{AB}^{\imath t} \rho_{AB}^{-\imath t} = \sigma_A^{\imath t}
\rho_{A}^{-\imath t} \otimes {\mathrm{I}}.
\enq
Since $f(t) = -\ln(t)$ is a non-affine and diffused operator convex function,
Eq. (\ref{f-eq}) should be satified
if Eq. (\ref{petz-eq}) is true, which, indeed, is the case.
Ruskai \cite{ruskai-equality-2002} gave the following conditions for the equality in
Eq. (\ref{mono}) for $f(t) = -\ln(t)$ as
\beq
\label{ruskai-eq}
\ln(\sigma_{AB}) - \ln(\rho_{AB}) = \left[ \ln(\sigma_A) - \ln(\rho_A) \right]
\otimes {\mathrm{I}}.
\enq
Ruskai showed that Eq. (\ref{ruskai-eq}) can be obtained from Eq. (\ref{petz-eq})
by taking the derivative of both sides of Eq. (\ref{petz-eq}) w.r.t. $t$ at $t=0$
\cite{ruskai-equality-2002}.

\begin{corollary}
For $f(t) = -\ln(t)$, the necessary and sufficient conditions for the equality in
Eq. (\ref{equality0}) are given by Eq. (\ref{f-eq}).
\end{corollary}
The proof is along the same lines as that of Lemma \ref{lemma-equality}.
It is interesting to note that following Ruskai's approach \cite{ruskai-equality-2002},
by taking the derivative of both sides of Eq. (\ref{f-eq}) w.r.t. $t$ at $t=0$, we obtain
$S_{(-\ln)}(\rho_{AB} || \sigma_{AB}) = $ $S_{(-\ln)}(\rho_{A} || \sigma_{A})$.

We now consider the following special case, which is applicable to a variety of cases for both
the \frel and the \fentns.
\begin{corollary}
\label{corollary-equality}
For a non-affine and diffused operator convex function $f(\cdot)$,
strictly positive $\rho_{AB}$, $\sigma_B$,
and $\rho_A = \tr_B(\rho_{AB})$, the equality in
\beq
\label{equality1}
S_f \left( \rho_{AB} || {\mathrm{I}} \otimes \sigma_B \right) \geq
S_f \left[\rho_A || \tr(\sigma_B) {\mathrm{I}} \right]
\enq
holds if and only if $\rho_{AB} = \rho_A \otimes \sigma_B/\tr(\sigma_B)$.
\end{corollary}
\begin{proof}
The inequality in Eq. (\ref{equality1}) is true, of course, because of monotonicity.
Using Eq. (\ref{dummy51}), the equality conditions are
\beqa
\label{dummy5-1}
\tr\left[ \left({\mathrm{I}} \otimes \phi_B^n \right) \rho_{AB}^{-n+1} \right]
& = & \tr\left(\rho_{A}^{-n+1} \right),
~~ \forall ~ n \geq 0,
\enqa
where $\phi_B = \sigma_B/\tr(\sigma_B)$.
Let the spectral decompositions of the matrices in the above equation be given by
$\rho_A = \sum_{j=1}^{d_\rho} \lambda_j \ket{j_A} \bra{j_A}$,
$\phi_B = $ $\sum_{i=1}^{d_\sigma} \beta_i \ket{i_B} \bra{i_B}$,
$\rho_{AB} = \sum_{k=1}^{d_\rho d_\sigma}
\alpha_k \ket{k_{AB}} \bra{k_{AB}}$, where $d_\rho$, $d_\sigma$ are the dimensions of
the Hilbert spaces describing $\rho_A$, $\sigma_B$ respectively.
Then Eq. (\ref{dummy5-1}) can be restated as
\beqa
\label{dummy5}
\sum_{j =1}^{d_\rho}
\sum_{i=1}^{d_\sigma} \sum_{k=1}^{d_\rho d_\sigma}
\left(\frac{\beta_i}{\alpha_k}\right)^n \alpha_k
\left|  \bra{j_A} \braket{i_B}{k_{AB}} \right|^2 & = &
\sum_{j =1}^{d_\rho} \lambda_j^{-n} \lambda_j, ~~ \forall ~ n \geq 0, \\
\sum_{j =1}^{d_\rho}
\sum_{i=1}^{d_\sigma} \sum_{k=1}^{d_\rho d_\sigma}
\left[ \left(\frac{\beta_i}{\alpha_k}\right)^n \alpha_k -
\left(\frac{1}{\lambda_j}\right)^n \beta_i \lambda_j \right]
\left|  \bra{j_A} \braket{i_B}{k_{AB}} \right|^2 & = & 0, ~~ \forall ~ n \geq 0.
\enqa
Let there be $M$ distinct eigenvalues of $\rho_A$ denoted by $\lambda^{(m)}$,
$m = 1,...,M$. We follow similar reasoning as in Lemma
\ref{lemma-equality} to claim that we have $M$ disjoint sets
${\mathcal{Q}}_m$, $m=1,...,M$, such that $\forall$ $(j,i,k) \in$
${\mathcal{Q}}_m$,
\beq
\frac{\alpha_k}{\beta_i} = \lambda_j = \lambda^{(m)}.
\enq
For completeness, we shall
also define ${\mathcal{Q}}_0 = \bigcap_{m=1}^M {\mathcal{Q}}_m^c$, where
${\mathcal{Q}}^c$ denotes the complement of ${\mathcal{Q}}$. It follows that
for $(j,i,k) \in {\mathcal{Q}}_0$, $\left|  \bra{j_A} \braket{i_B}{k_{AB}} \right|=0$.
We now have
\beqa
\tr \left[ \rho_{AB} \left( \rho_A \otimes \phi_B \right) \right] & = &
\sum_{j=1}^{d_\rho} \sum_{i=1}^{d_\sigma} \sum_{k=1}^{d_\rho d_\sigma}
\alpha_k \lambda_j \beta_i \left|  \bra{j_A} \braket{i_B}{k_{AB}} \right|^2 \\
& = & \sum_{m=1}^M \sum_{(j,i,k) \in {\mathcal{Q}}_m}
\alpha_k \lambda_j \beta_i \left|  \bra{j_A} \braket{i_B}{k_{AB}} \right|^2 \\
& = & \sum_{m=1}^M \sum_{(j,i,k) \in {\mathcal{Q}}_m}
\alpha_k^2 \left|  \bra{j_A} \braket{i_B}{k_{AB}} \right|^2 \\
& = & \sum_{j=1}^{d_\rho} \sum_{i=1}^{d_\sigma} \sum_{k=1}^{d_\rho d_\sigma}
\alpha_k^2 \left|  \bra{j_A} \braket{i_B}{k_{AB}} \right|^2 \\
& = & \tr\left(\rho_{AB}^2 \right).
\enqa
Similarly, one can show that
\beqa
\tr \left( \rho_A^2 \otimes \phi_B^2 \right) & = &
\tr\left(\rho_{AB}^2 \right).
\enqa
Using the above two equations, it now follows that
\beqa
||\rho_{AB} - \rho_A \otimes \phi_B||_{\mathrm{F}}^2 & = &
\tr\left(\rho_{AB}^2 \right) + \tr \left( \rho_A^2 \otimes \phi_B^2 \right)
- 2 \tr \left[ \rho_{AB} \left( \rho_A \otimes \phi_B \right) \right] \\
& = & 0,
\enqa
or
$\rho_{AB} = \rho_A \otimes \sigma_B/\tr(\sigma_B)$.
\end{proof}

\begin{corollary}
The inequality
\beq
S_f \left( \rho_{AB} || {\mathrm{I}} \otimes \rho_B \right) \leq
S_f \left( \rho_{ABC} || {\mathrm{I}} \otimes \rho_{BC} \right)
\enq
holds and the equality conditions are given by
\beq
\tr\left[ \left( {\mathrm{I}} \otimes \rho_B ^{\imath t} \right)
\rho_{AB}^{-\imath t+1} \right] =
\tr\left[ \left( {\mathrm{I}} \otimes \rho_{BC}^{\imath t} \right)
\rho_{ABC}^{-\imath t+1} \right], ~~ \forall ~ t \in \mathbb{C}.
\enq
\end{corollary}
Note that the inequality follows immediately from the monotonicity and the equality
conditions from Eq. (\ref{f-eq}). Ruskai
\cite{ruskai-equality-2002} showed that the above inequality for $f(t) = -\ln(t)$
is just a restatement of the strong sub-additivity in Eq. (\ref{ssa}).

It follows that for strictly positive $\rho$ and $\sigma$, and a non-affine and diffused operator
convex function $f(\cdot)$, that
\beq
S_f(\rho || \sigma) \geq f\left[ \frac{\tr(\sigma)}{\tr(\rho)} \right] \tr(\rho)
\enq
with equality if and only if $\rho/\tr(\rho) = \sigma/\tr(\sigma)$.
To see this, substitute $\rho_{AB} = \rho$,
$\rho_A = \tr(\rho)$, and $\sigma_B = \sigma$ in Eq. (\ref{equality1}).

However, Petz showed the same result for a non-affine operator convex function with
no requirement that the function has to be diffused \cite{petz-quasi-entr-1986}.
We provide an alternate derivation of his result.

\begin{lemma}
(Petz \cite{petz-quasi-entr-1986})
For strictly positive $\rho$ and $\sigma$, and a non-affine operator convex function
$f(\cdot)$, the following holds
\beq
\label{frellb}
S_f(\rho || \sigma) \geq f\left[ \frac{\tr(\sigma)}{\tr(\rho)} \right] \tr(\rho)
\enq
with equality if and only if $\rho/\tr(\rho) = \sigma/\tr(\sigma)$.
\end{lemma}
\begin{proof}
We first note that since $f(\cdot)$ is a non-affine operator convex function, hence,
it is strictly convex. Secondly, note that for a
strictly convex function $f(\cdot)$, $a_k > 0$, $k=1,...,d$,
\beq
\label{fsum-inq}
\sum_{k=1}^d a_k f \left( \frac{b_k}{a_k} \right) \geq
\left( \sum_{k=1}^d a_k \right)
f \left( \frac{\sum_{k=1}^d b_k}{\sum_{k=1}^d a_k} \right),
\enq
with equality if and only if $b_k/a_k$ is constant $\forall$ $k=1,...,d$.
Let the spectral decompositions of $\rho$ and $\sigma$
be given by Eqs. (\ref{sdrho}) and (\ref{sdsigma}) respectively.
Using Eqs. (\ref{frentropy4}) and (\ref{fsum-inq}), we get
\beqa
S_f(\rho || \sigma) & = & \sum_{i,j}^d p_i | \braket{i_\sigma}{j_\rho}|^2
f \left( \frac{q_j}{p_i} \right) \\
& \geq & \sum_{j=1}^d  \left( \sum_{i=1}^d p_i | \braket{i_\sigma}{j_\rho}|^2
\right) f \left( \frac{\sum_{i=1}^d q_j | \braket{i_\sigma}{j_\rho}|^2}
{\sum_{i=1}^d p_i | \braket{i_\sigma}{j_\rho}|^2} \right) \\
& \geq &  \left( \sum_{i,j=1}^d p_i | \braket{i_\sigma}{j_\rho}|^2
\right) f \left( \frac{\sum_{i,j=1}^d q_j | \braket{i_\sigma}{j_\rho}|^2}
{\sum_{i,j=1}^d p_i | \braket{i_\sigma}{j_\rho}|^2} \right) \\
& = & f\left[ \frac{\tr(\sigma)}{\tr(\rho)} \right] \tr(\rho).
\enqa
The conditions for the equality are
$q_j | \braket{i_\sigma}{j_\rho}|^2 = $ $c_1 p_i
| \braket{i_\sigma}{j_\rho}|^2$, $\forall$ $i=1,...,d$, and
$q_j = $
$c_2 \sum_{i=1}^d p_i | \braket{i_\sigma}{j_\rho}|^2$, $\forall$
$j=1,...,d$, where $c_1$, $c_2$ are positive constants. It now follows that
$c_1 = c_2$ $= \tr(\sigma)/\tr(\rho)$, $\tr(\rho \sigma) = \tr(\sigma^2)/c_1$,
and $\tr(\rho^2) = \tr(\sigma^2)/c_1^2$.
Let $||\kappa||_{\mathrm{F}}$ $=\sqrt{\tr\left(\kappa^\dagger \kappa\right)}$ denote the
Frobenius norm of $\kappa$. Then
\beq
\left| \left| \frac{\rho}{\tr(\rho)} - \frac{\sigma}{\tr(\sigma)} \right|\right|_{\mathrm{F}}^2
= \frac{\tr(\rho^2)}{\left[ \tr(\rho) \right]^2} +
\frac{\tr(\sigma^2)}{\left[ \tr(\sigma) \right]^2} - 2
\frac{\tr(\rho \sigma)}{\tr(\rho) \tr(\sigma)} = 0.
\enq
QED.
\end{proof}

The Klein's inequality is a special case of the above Lemma for $f(t) = -\ln(t)$.

\subsection{Convexity}
\label{sub-convexity}

We now examine the convexity properties of the \frel.

\begin{lemma}
\label{lemma-g}
Let
\beq
\label{eqg}
g(x) \define \sqrt{x} f\left( \frac{1}{x} \right) \sqrt{x} = x f\left( \frac{1}{x} \right),
~~ x \in (0,\infty).
\enq
Then if $f(\cdot)$ is operator convex, so is $g(\cdot)$.
\end{lemma}
\begin{proof}
The proof is not much different from that of Theorem 2.2 in Ref. \cite{effros-2009}
though without using the linear super-operators.
Note first from Lemma 5.1.5 in Ref. \cite{bhatia-1997} that if $A \leq B$, then for
any matrix $X$ with appropriate dimensions, $X^\dagger A X \leq X^\dagger B X$.

For $0 \leq \lambda \leq 1$, choose $n=2$, strictly positive $A$, $B$,
$C = \lambda A + (1-\lambda) B$, $E_1 = \sqrt{\lambda AC^{-1}}$,
$E_2 = \sqrt{(1-\lambda) B C^{-1}}$, $\phi_1 = A^{-1}$, and $\phi_2 = B^{-1}$,
and substitute in Eq. (\ref{jensen}) to get
\beq
f \left( C^{-1} \right) \leq \sqrt{C^{-1}} \left[
\lambda \sqrt{A} f\left( A^{-1} \right)
\sqrt{A} + (1-\lambda) \sqrt{B} f\left( B^{-1} \right) \sqrt{B} \right] \sqrt{C^{-1}}.
\enq
The results follows by pre-multiplying and
post-multiplying both sides by $\sqrt{C}$ and noting that
\beq
g(X) = \sqrt{X} f \left( X^{-1} \right) \sqrt{X}.
\enq
QED.
\end{proof}
\ \\ \\
The operator convexity of $g(t) = t \ln(t)$ follows from that $f(t) = -\ln(t)$
by using the above result.
It is easy to see that from Eq. (\ref{frentropy1}) that the \frel is convex in the
second argument since
\beqa
S_f\left[ \rho || \lambda \sigma_1 + (1-\lambda) \sigma_2 \right] & = &
\nvec{\sqrt{\rho}}^\dagger
f \left\{ \left[\lambda \sigma_1 + (1-\lambda) \sigma_2\right]
\otimes \left(\rho^{-1}\right)^\transpose \right\} \nvec{\sqrt{\rho}} \\
& \leq &
\lambda \nvec{\sqrt{\rho}}^\dagger
f \left[ \sigma_1
\otimes (\rho^{-1})^\transpose \right] \nvec{\sqrt{\rho}} + \nonumber \\
& & ~~~~ (1-\lambda) \nvec{\sqrt{\rho}}^\dagger
f \left[\sigma_2
\otimes \left(\rho^{-1}\right)^\transpose \right] \nvec{\sqrt{\rho}} \\
\label{sec_convexity}
& = & \lambda S_f(\rho || \sigma_1) + (1-\lambda) S_f(\rho || \sigma_2).
\enqa
It is easy to check from Eq. (\ref{frentropy4}) that
\beq
S_f(\rho || \sigma) = S_g(\sigma || \rho),
\enq
where $g(\cdot)$ is as defined in Eq. (\ref{eqg}).
It now follows from Eq. (\ref{sec_convexity}) that the \frel is convex in its first argument
as well since
\beqa
S_f\left[\lambda \rho_1 + (1-\lambda) \rho_2 || \sigma \right] & = &
S_g \left[ \sigma || \lambda \rho_1 + (1-\lambda) \rho_2 \right] \\
& \leq & \lambda S_g(\sigma || \rho_1) + (1-\lambda) S_g(\sigma || \rho_2) \\
& = & \lambda S_f(\rho_1 || \sigma) + (1-\lambda) S_f(\rho_2 || \sigma).
\enqa

We now show that the \frel is jointly convex in its arguments which is a stronger
result than the convexity of any one of its arguments. Petz proved the joint
convexity of quantum quasi relative entropy \cite{petz-quasi-entr-1986}. We provide
an alternate proof that is more accessible. Furthermore, we give
the equality conditions for a class of operator convex functions.
\begin{lemma}
For $0 < \lambda < 1$, strictly positive $\rho_1$, $\rho_2$, $\sigma_1$,
$\sigma_2$, and $f(\cdot)$ operator convex,
\beq
\label{frel-convex}
S_f\left( \rho_\lambda || \sigma_\lambda \right) \leq \lambda S_f(\rho_1 || \sigma_1)
+ (1-\lambda) S_f(\rho_2 || \sigma_2),
\enq
where $\rho_\lambda = \lambda \rho_1 + (1-\lambda) \rho_2$ and
$\sigma_\lambda = \lambda \sigma_1 + (1-\lambda) \sigma_2$. The equality holds
for a non-affine and diffused operator convex function $f(\cdot)$
if and only if
\beq
\label{convexity-equality}
\tr \left( \sigma_\lambda^{\imath t} \rho_\lambda^{-\imath t+1} \right) =
\lambda \tr \left( \sigma_1^{\imath t} \rho_1^{-\imath t+1} \right) +
(1 - \lambda) \tr \left( \sigma_2^{\imath t} \rho_2^{-\imath t+1} \right), ~~~
\forall ~ t \in {\mathbb{C}}.
\enq
\end{lemma}
\begin{proof}
Choose
\beqa
E_1 & = & \sqrt{\lambda} \left[{\mathrm{I}} \otimes
\left( \sqrt{\rho_1}\right)^\transpose\right]
\left[{\mathrm{I}} \otimes \left(\rho_\lambda^{-1/2}\right)^\transpose \right], \\
E_2 & = & \sqrt{1-\lambda} \left[{\mathrm{I}} \otimes
\left[ \sqrt{\rho_2}\right)^\transpose\right]
\left[{\mathrm{I}} \otimes \left(\rho_\lambda^{-1/2}\right)^\transpose \right], \\
\phi_1 & = & \sigma_1 \otimes \left(\rho_1^{-1}\right)^\transpose, \\
\phi_2 & = & \sigma_2 \otimes \left(\rho_2^{-1}\right)^\transpose.
\enqa
It is easy
to check that $E_1^\dagger E_1 + E_2^\dagger E_2 = {\mathrm{I}}$ and
$E_1^\dagger \phi_1 E_1 + E_2^\dagger \phi_2 E_2 = \sigma_\lambda \otimes
\left(\rho_\lambda^{-1} \right)^\transpose$. Using Eq. (\ref{jensen}),
and pre-multiplying both sides by
$X^\dagger = \left[{\mathrm{I}} \otimes \left(\sqrt{\rho_\lambda}\right)^* \right]$
and post-multiplying both sides by
$X = \left[{\mathrm{I}} \otimes \left(\sqrt{\rho_\lambda}\right)^\transpose \right]$,
we get
\beqa
X^\dagger
f \left[ \sigma_\lambda \otimes \left(\rho_\lambda^{-1} \right)^\transpose \right] X
& \leq & \lambda \left[{\mathrm{I}} \otimes \left( \sqrt{\rho_1}\right)^*\right]
f \left[ \sigma_1 \otimes \left(\rho_1^{-1}\right)^\transpose \right]
\left[{\mathrm{I}} \otimes \left(\sqrt{\rho_1}\right)^\transpose \right] \nonumber \\
& & ~~~ + (1-\lambda)
\left[{\mathrm{I}} \otimes \left( \sqrt{\rho_2}\right)^*\right]
f \left[ \sigma_2 \otimes \left(\rho_2^{-1}\right)^\transpose \right]
\left[{\mathrm{I}} \otimes \left(\sqrt{\rho_2}\right)^\transpose \right].
\enqa
Pre-multiplying both sides by $\nvec{{\mathrm{I}}}^\dagger$, post-multiplying
by $\nvec{{\mathrm{I}}}$, and using Eq. (\ref{frentropy4}), we get
\beq
S_f(\rho_\lambda || \sigma_\lambda) \leq \lambda S(\rho_1 || \lambda_1) 
+ (1-\lambda) S(\rho_2 || \lambda_2).
\enq
To prove the equality conditions, we follow the analysis in Lemma \ref{lemma-equality} to
reduce the equality conditions to
\beq
\tr \left( \sigma_\lambda^n \rho_\lambda^{-n+1}\right)
= \lambda \tr \left( \sigma_1^n \rho_1^{-n+1}\right) +
(1- \lambda) \tr \left( \sigma_2^n \rho_2^{-n+1}\right), ~~ \forall ~ n \geq 0,
\enq
which, using the reasoning in Lemma \ref{lemma-equality},
can be shown to be equivalent to Eq. (\ref{convexity-equality}).
\end{proof}

For $f(t) = -\ln(t)$, Ruskai \cite{ruskai-equality-2002} gave the equality
conditions for Eq. (\ref{frel-convex}) as
\beq
\ln(\sigma_\lambda) - \ln(\rho_\lambda) = \ln(\sigma_i) - \ln(\rho_i), ~~ i = 1,2.
\enq

\begin{corollary}
The \frel is sub-additive, i.e., for strictly positive $\rho_i$, $\sigma_i$, $i=1,2$,
\beq
S_f\left( \rho_1 + \rho_2 || \sigma_1 + \sigma_2 \right) \leq
S_f \left( \rho_1 || \sigma_1 \right) + S_f \left(\rho_2 || \sigma_2 \right)
\enq
and the equality holds if and only if
\beq
\tr \left[ \left(\sigma_1+\sigma_2\right)^{\imath t}
\left(\rho_1+\rho_2\right)^{-\imath t+1} \right] =
\tr \left( \sigma_1^{\imath t} \rho_1^{-\imath t+1} \right)
+ \tr \left( \sigma_2^{\imath t} \rho_2^{-\imath t+1} \right),
~~ \forall ~ t \in {\mathbb{C}}.
\enq
\end{corollary}
\begin{proof}
Joint convexity implies the sub-additivity of the \frel since
\beqa
S_f\left( \rho_1 + \rho_2 || \sigma_1 + \sigma_2 \right)
& = & S_f\left( \frac{2 \rho_1}{2} + \frac{2 \rho_2}{2} \Big|\Big| \frac{2 \sigma_1}{2}
+ \frac{2 \sigma_2}{2} \right) \\
& \leq & \frac{1}{2} S_f \left( 2 \rho_1 || 2 \sigma_1 \right) +
\frac{1}{2} S_f \left( 2 \rho_2 || 2 \sigma_2 \right) \\
& = & S_f \left( \rho_1 || \sigma_1 \right) + S_f \left(\rho_2 || \sigma_2 \right).
\enqa
The equality conditions follow from Eq. (\ref{convexity-equality}).
\end{proof}

\begin{lemma}
Joint convexity of the \frel implies monotonicity, and for a completely positive
trace-preserving (CPTP) quantum operation ${\mathcal E}(\cdot)$,
\beq
\label{mono_op}
S_f\left[ {\mathcal E}(\rho) || {\mathcal E}(\sigma) \right]
\leq S_f(\rho||\sigma),
\enq
and the equality holds if and only if
\beq
\tr \left\{ \left[{\mathcal E}(\sigma)\right]^{\imath t}
\left[{\mathcal E}(\rho) \right]^{-\imath t+1} \right\} =
\tr \left( \sigma^{\imath t} \rho^{-\imath t+1} \right), ~~ \forall ~ t \in {\mathbb{C}}.
\enq
\end{lemma}
We omit the proof.

\section{Quantum $f$-entropy}

We now define \fent for strictly positive $\rho$, denoted by $S_f(\rho)$, in terms of the \frel as
\beq
S_f(\rho) \define - S_f\left( \rho || {\mathrm{I}} \right) =
-\tr\left[ \rho f\left( \rho^{-1} \right) \right] = -\sum_{i=1}^d p_i
f\left( \frac{1}{p_i} \right),
\enq
where the spectral decomposition of $\rho$ is given by
\beq
\rho = \sum_{i=1}^d p_i \ket{i_\rho} \bra{i_\rho},
\enq
and $d$ is the dimension of the Hilbert space that describes $\rho$.
For a density matrix $\rho$, and $f(t) = -\ln(t)$, \fent coincides with the
von-Neumann entropy \cite{petz-book,nielsen-chuang}.
\begin{lemma}
\label{ent-lemma1}
The following holds:
\bit
\item[(i)] For strictly positive $\rho$ with dimension $d$,
\beq
S_f(\rho) \leq -\tr(\rho) f\left[ \frac{d}{\tr{(\rho})}\right].
\enq
For a non-affine operator convex function $f(\cdot)$, the equality holds
if and only if $\rho = \tr(\rho) {\mathrm{I}}/d$.
\item[(ii)] Let the joint state in system $AB$ be a pure state. Then $S(A) = S(B)$.
\item[(iii)] Projective measurements increase \fentns, and for a non-affine and diffused
operator convex function $f(\cdot)$, the equality holds if and only if the projective
measurement leaves the state unchanged.
\eit 
\end{lemma}
\begin{proof}
\bit
\item[(i)] follows by using Eq. (\ref{frellb}).
\item[(ii)] Let the joint state in system $AB$ be a pure state denoted by
$\ket{\phi_{AB}}$, and let its Schmidt decomposition be given by
\beq
\ket{\phi_{AB}} = \sum_k \sqrt{\lambda_k} \ket{k_A} \ket{k_B}
\enq
where $\{\lambda_k\}$ is a probability vector,
$\{\ket{k_A}\}$ and $\{\ket{k_B}\}$ are orthonormal states in $A$
and $B$ respectively. Then since
\beqa
\rho_A = \tr_B \left(\ket{\phi_{AB}} \bra{\phi_{AB}} \right) & = & \sum_k \lambda_k
\ket{k_A} \bra{k_A}, \\
\rho_B = \tr_A \left(\ket{\phi_{AB}} \bra{\phi_{AB}} \right) & = & \sum_k \lambda_k
\ket{k_B} \bra{k_B},
\enqa
it follows that
\beq
S_f \left(\rho_A\right) = S_f \left(\rho_B\right).
\enq
\item[(iii)] Let $\{P_i\}$ be a complete set of projectors and
$\sum_i P_i = {\mathrm{I}}$. Projective measurements increase \fent since
it follows using Eq. (\ref{mono_op}) that
\beqa
S_f(\rho) & = & -S_f(\rho || {\mathrm{I}}) \\
& \geq & -S_f \left( \sum_i P_i \rho P_i \Big| \Big| {\mathrm{I}} \right) \\
& = & S_f\left(\sum_i P_i \rho P_i \right).
\enqa
\eit
To prove the equality condition, we use Eq. (\ref{mono_op}) to get
\beq
\tr\left(\rho^{-n+1}\right) = \tr \left[ \left(\sum_i P_i \rho P_i \right)^{-n+1} \right],
~~ \forall ~ n \geq 0,
\enq
or the eigenvalues of $\rho$ and $\sum_i P_i \rho P_i$ are the same including
multiplicities. In particular,
\beq
\tr(\rho^2) = \tr\left[\left( \sum_i P_i \rho P_i \right)^2 \right].
\enq
Then it is easy to show that
\beq
\Big|\Big|\rho - \sum_i P_i \rho P_i \Big|\Big|_{\mathrm{F}}^2 = \tr(\rho^2) -
\tr\left[\left( \sum_i P_i \rho P_i \right)^2 \right] = 0,
\enq
which proves the result.
\end{proof}

\begin{lemma}
Let $\{\rho_i\}$ be a set of strictly positive matrices with unit trace and
described by the same
Hilbert space, and let their spectral decompositions be given by
\beq
\label{sdrho_i}
\rho_i = \sum_j q_{ij} \ket{i,j} \bra{i,j}
\enq
Then for any probability vector $\{p_i\}$,
\beq
\sum_i p_i S_f(\rho_i) \leq S_f\left( \sum_i p_i \rho_i \right) \leq
-\sum_{ij} p_i q_{ij} f\left( \frac{1}{p_i q_{ij}} \right).
\enq
For a non-affine and diffused operator convex $f(\cdot)$,
the equality in the first inequality holds if and
only if the $\rho_i$'s with $p_i>0$ are identical, and the equality in the second inequality
holds if and only $\rho_i$'s have support on orthogonal subspaces.
\end{lemma}
\begin{proof}
Consider a joint state in system $AB$ as
\beq
\rho_{AB} = \sum_i p_i \rho_i \otimes \ket{i_B} \bra{i_B},
\enq
where $\rho_i$ lie in the system $A$ (our system of interest),
and $\{\ket{i_B}\}$ is any orthonormal basis in system $B$ (ancilla).
Then for
\beq
\rho_B = \tr_A(\rho_{AB}) = \sum_i p_i \ket{i_B} \bra{i_B},
\enq
we have
\beqa
S_f\left( \rho_{AB} || {\mathrm{I}} \otimes \rho_B \right) & = &
S_f \left(\sum_{i,j} p_i q_{ij} \ket{i,j} \bra{i,j} \otimes \ket{i_B} \bra{i_B} ~ \Big|\Big| ~
\sum_{i,j} p_i  \ket{i,j} \bra{i,j} \otimes \ket{i_B} \bra{i_B} \right) ~~~~ \\
& = & \sum_{i,j} f\left( \frac{1}{q_{ij}} \right) p_i q_{ij} \\
& = & -\sum_i p_i S_f(\rho_i).
\enqa
Note that
\beq
\rho_A = \tr_B\left(\rho_{AB}\right) = \sum_i p_i \rho_i.
\enq
It now follows that
\beqa
S_f\left( \sum_i p_i \rho_i \right) & = & -S_f \left( \sum_i p_i \rho_i \Big| \Big| {\mathrm{I}}
\right) \\
& = & -S_f \left( \rho_A || {\mathrm{I}} \right) \\
& \geq & -S_f \left( \rho_{AB} || {\mathrm{I}} \otimes \rho_B \right) \\
& = & \sum_i p_i S_f(\rho_i).
\enqa
Using Eq. (\ref{equality1}), the equality holds if and only if
$\rho_{AB} = \rho_A \otimes \rho_B$, or
\beq
\sum_i p_i \rho_i \otimes \ket{i_B} \bra{i_B} = \sum_i p_i \rho_i \otimes
\sum_k p_k \ket{k_B} \bra{k_B},
\enq
or the $\rho_i$ with $p_i > 0$ are identical.

To prove the upper bound, let us first assume that $\rho_i$'s are all pure
and $\rho_i = \ket{\psi_i} \bra{\psi_i}$. We attach an ancilla $B$ to our system $A$
such that
\beq
\ket{\phi_{AB}} = \sum_i \sqrt{p_i} \ket{\psi_i} \ket{i_B},
\enq
where $\{ \ket{i_B} \}$ is an orthonormal basis in $B$. It easily follows that
\beqa
\rho_A & = & \tr_B \left( \ket{\phi_{AB}} \bra{\phi_{AB}} \right) = \sum_i p_i \rho_i, \\
\rho_B & = & \tr_A \left( \ket{\phi_{AB}} \bra{\phi_{AB}} \right) =
\sum_i \sqrt{p_i p_j} \braket{\psi_j}{\psi_i} \ket{i_B} \bra{j_B}.
\enqa
Define projectors $P_i = \ket{i_B} \bra{i_B}$ and
\beq
\rho_{\acute{B}} = \sum_i P_i \rho_B P_i = \sum_i p_i \ket{i_B} \bra{i_B}.
\enq
Hence using Lemma \ref{ent-lemma1}, we have
\beqa
-\sum_i p_i f \left( \frac{1}{p_i} \right) = S_f \left( \rho_{\acute{B}} \right)
& \geq & S_f \left( \rho_B \right) = S_f \left( \rho_A \right) \\
& = & S_f\left( \sum_i p_i \rho_i \right) = S_f\left( \sum_i p_i \ket{\psi_i} \bra{\psi_i}\right).
\enqa
Using Lemma \ref{ent-lemma1},
the equality holds if and only if $\rho_B = \rho_{\acute{B}}$ or
$\braket{\psi_j}{\psi_i} = \delta_{i,j}$. This proves the Lemma when $\rho_i$'s
are pure. For mixed $\rho_i$'s, we use the above result to have
\beq
S_f \left( \sum_{i,j} p_i q_{i,j} \ket{i,j} \bra{i,j} \right) \leq
- \sum_{i,j} p_i q_{i,j}  f\left( \frac{1}{p_i q_{ij}} \right),
\enq
with equality if and only if $\rho_i$ have support on orthogonal subspaces.
\end{proof}

\section{Generalized Data Processing Inequalities}

In this section, we show that the $f$-generalizations of well-known quantum
information theoretic quantities also satisfy the data processing inequalities
\cite{nielsen-chuang,covertom} as they do for $f(t) = -\ln(t)$.

\subsection{Holevo information}

Consider a state in the composite system consisting of $A$ and $B$ given by
\beq
\label{dummy6}
\rho_{AB} = \sum_i p_i \rho_i \otimes \ket{i_B} \bra{i_B},
\enq
where $\{\ket{i_B}\}$ is an orthonormal basis in $B$ and
$\rho_i$'s are strictly positive with unit trace.. Let ${\mathcal E}(\cdot)$
denote the CPTP quantum operation acting on $A$. Then $f$-Holevo
$\chi_f({\mathcal E})$ quantity is defined as
\beq
\chi_f({\mathcal E}) = \max_{\{p_i, \rho_i\}}
S_f \left[ \left({\mathcal E} \otimes {\mathcal I}_B \right) \rho_{AB} ||
{\mathcal E} (\rho_A) \otimes \rho_B \right],
\enq
where $\rho_{AB}$ is given by Eq. (\ref{dummy6}), $\rho_A = \tr_B(\rho_{AB})$,
and $\rho_B = \tr_A(\rho_{AB})$. For $f(\cdot) = -\ln(\cdot)$, this
quantity is the product state capacity for the quantum
channel ${\mathcal{E}}(\cdot)$ for transmitting classical information
as proved by Holevo, Schumacher, and Westmoreland (HSW theorem)
\cite{holevo-hsw-1998, schu-west-hsw-1997}.

We note here that $\chi_f({\mathcal E})$ is independent of the choice of
$\{\ket{i_B}\}$. To see this, let $\{U\ket{i_B}\}$ be the new orthonormal basis chosen
for the system $B$ associated with
$\acute{\chi}_f({\mathcal E})$. Then for $V = {\mathrm{I}} \otimes U$, we have
\beqa
\acute{\chi}_f({\mathcal E}) & = & \max_{\{p_i, \rho_i\}}
S_f \Bigg\{ V \left[ \sum_i p_i {\mathcal E}(\rho_i)
\otimes \ket{i_B} \bra{i_B} \right] V^\dagger \Big| \Big| \\
& & ~~~~~~~~~~~~~~~~~~~~~~~~~~
V \left[ \sum_i p_i {\mathcal E}(\rho_i)
\otimes \sum_j p_j \ket{j_B} \bra{j_B} \right] V^\dagger \Bigg\} \\
& = & \chi_f({\mathcal E}).
\enqa
Let ${\mathcal E}_1(\cdot)$ and ${\mathcal E}_2(\cdot)$ be two CPTP quantum
operations, then using Eq. (\ref{mono_op}), we have
\beqa
\chi_f({\mathcal E}_1) & = & \max_{\{p_i, \rho_i\}}
S_f \left[ \left({\mathcal E}_1 \otimes {\mathcal I}_B \right) \rho_{AB} ||
\left({\mathcal E}_1 \otimes {\mathcal I}_B \right) (\rho_A \otimes \rho_B) \right] \\
& \geq & \max_{\{p_i, \rho_i\}}
S_f \left[ \left({\mathcal E}_2 \otimes {\mathcal I}_B \right)
\left({\mathcal E}_1 \otimes {\mathcal I}_B \right) \rho_{AB} ||
\left({\mathcal E}_2 \otimes {\mathcal I}_B \right)
\left({\mathcal E}_1 \otimes {\mathcal I}_B \right) (\rho_A \otimes \rho_B) \right] \\
& = & \chi_f({\mathcal E}_2 \circ {\mathcal E}_1).
\enqa
\beqa
\chi_f({\mathcal E}_2 \circ {\mathcal E}_1) & = & \max_{\{p_i, \rho_i\}}
S_f \Bigg\{ \left({\mathcal E}_2 \otimes {\mathcal I}_B \right)
\left[ \sum_i p_i {\mathcal E}_1(\rho_i) \otimes \ket{i_B} \bra{i_B} \right] \Big|\Big|
\nonumber \\
& & ~~~~~~~~~~~~~~
\left({\mathcal E}_2 \otimes {\mathcal I}_B \right)
\left[\sum_i p_i {\mathcal E}_1(\rho_i) \otimes \rho_B\right] \Bigg\}  \\
& = & \max_{\{p_i, \sigma_i: \sigma_i = {\mathcal{E}_1}(\rho_i)\}}
S_f \Bigg\{ \left({\mathcal E}_2 \otimes {\mathcal I}_B \right)
\left[ \sum_i p_i \sigma_i \otimes \ket{i_B} \bra{i_B} \right] \Big|\Big| \nonumber \\
& & ~~~~~~~~~~~~~~
\left({\mathcal E}_2 \otimes {\mathcal I}_B \right)
\left[\sum_i p_i \sigma_i \otimes \rho_B\right] \Bigg\}  ~~~~~~~~ \\
& \leq & \chi_f({\mathcal E}_2).
\enqa
Hence,
\beq
\chi_f({\mathcal E}_2 \circ {\mathcal E}_1) \leq
\min\left\{ \chi_f({\mathcal E}_1), \chi_f({\mathcal E}_2) \right\},
\enq
which is the data processing inequality.

\subsection{Entanglement-assisted capacity}

Bennett \emph{et al} gave an expression for the capacity known as the
entanglement-assisted classical capacity if the sender and receiver have 
a shared quantum entanglement \cite{ent-ass-cap-2002}.

Let $Q$ be the system of interest and the purification of a state in $Q$ is
given in the joint system $RQ$.
Then the $f$-generalization of the entanglement-assisted channel capacity is defined as 
\beq
C_{E,f}({\mathcal E}) = \max_{\rho_Q} S_f\left[
\left({\mathcal E} \otimes {\mathcal I}_R \right) \left( \ket{\psi_{QR}} \bra{\psi_{QR}}
\right) \Big| \Big| \left({\mathcal E} \otimes {\mathcal I}_R \right) (\rho_Q \otimes \rho_R)
\right],
\enq
where $\ket{\psi_{QR}}$ is a purification of the density matrix $\rho_Q$ and $\rho_R =
\tr_Q\left( \ket{\psi_{QR}} \bra{\psi_{QR}} \right)$.
\beqa
C_{E,f}({\mathcal E}_1) & = & \max_{\rho_Q} S_f\left[
\left({\mathcal E}_1 \otimes {\mathcal I}_R \right) \left( \ket{\psi_{QR}} \bra{\psi_{QR}}
\right) \Big| \Big| \left({\mathcal E}_1 \otimes {\mathcal I}_R \right)
(\rho_Q \otimes \rho_R) \right] \\
& \geq & \max_{\rho_Q} S_f\Big[ \left({\mathcal E}_2 \otimes {\mathcal I}_R \right)
\left({\mathcal E}_1 \otimes {\mathcal I}_R \right) \left( \ket{\psi_{QR}} \bra{\psi_{QR}}
\right) \Big| \Big| \\
& & ~~~~~~~~~~~~~~~~~~~~~~~~~~~~
\left({\mathcal E}_2 \otimes {\mathcal I}_R \right)
\left({\mathcal E}_1 \otimes {\mathcal I}_R \right) (\rho_Q \otimes \rho_R)
\Big] \\
& = & C_{E,f}({\mathcal E}_2 \circ {\mathcal E}_1).
\enqa

Let us introduce an ancilla $E_1$ with a Unitary operation $V_1$ over the composite
system $QRE_1$ to mock up the quantum operation ${\mathcal{E}_1} \otimes
{\mathcal{I}}_R$, i.e.,
\beq
\left({\mathcal{E}_1} \otimes {\mathcal{I}}_R\right) (\rho_{QR}) =
\tr_{E_1} \left[V_1 \left( \rho_{QR} \otimes
\ket{0_{E_1}} \bra{0_{E_1}} \right) V_1^\dagger \right],
\enq
where $\ket{0_{E_1}}$ is the initial state of the ancilla.
Let
$\ket{\psi_{QRE_1}} = \ket{\psi_{QR}} \ket{0_{E_1}}$ and
$\rho_{E_1} = \ket{0_{E_1}}\bra{0_{E_1}}$.
Then
\beqa
C_{E,f}({\mathcal E}_2 \circ {\mathcal E}_1) & = & \max_{\rho_Q} S_f\Big\{
({\mathcal E}_2 \otimes {\mathcal I}_R) \left[ \tr_{E_1} \left(V_1 \ket{\psi_{QRE_1}}
\bra{\psi_{QRE_1}} V_1^\dagger \right) \right] \Big| \Big| \nonumber \\
& & ~~~~~~~~~~~ {\mathcal E}_2 \left[\tr_{R E_1} \left(V_1 \ket{\psi_{QRE_1}}
\bra{\psi_{QRE_1}} V_1^\dagger \right)\right] \otimes \rho_R \Big\} \\
& = & \max_{\rho_Q} S_f\Big(
\tr_{E_1} \left[ \left({\mathcal E}_2 \otimes {\mathcal I}_R \otimes {\mathcal I}_{E_1}
\right) \left(V_1 \ket{\psi_{QRE_1}}
\bra{\psi_{QRE_1}} V_1^\dagger \right) \right] \Big| \Big| \nonumber \\
& & ~~~~~~~~~~~ \tr_{E_1} \left\{ {\mathcal E}_2 \left[\tr_{R E_1} \left(V_1
\ket{\psi_{QRE_1}} \bra{\psi_{QRE_1}} V_1^\dagger \right)\right] \otimes
\rho_R \otimes \rho_{E_1} \right\} \Big) \\
& \leq & \max_{\rho_Q} S_f\Big\{
\left({\mathcal E}_2 \otimes {\mathcal I}_R \otimes {\mathcal I}_{E_1}
\right) \left(V_1 \ket{\psi_{QRE_1}}
\bra{\psi_{QRE_1}} V_1^\dagger \right) \Big| \Big| \nonumber \\
& & ~~~~~~~~~~~ {\mathcal E}_2 \left[\tr_{R E_1} \left(V_1
\ket{\psi_{QRE_1}} \bra{\psi_{QRE_1}} V_1^\dagger \right)\right] \otimes
\rho_R \otimes \rho_{E_1} \Big\} \\
& = & \max_{{\mathcal{E}_1}(\rho_Q)} S\Big\{
\left({\mathcal E}_2 \otimes {\mathcal I}_R \otimes {\mathcal I}_{E_1}
\right) \left(\ket{\acute{\psi}_{QRE_1}}
\bra{\acute{\psi}_{QRE_1}} \right) \Big| \Big| \nonumber \\
& & ~~~~~~~~~~~ {\mathcal E}_2 \left[\tr_{R E_1} \left(
\ket{\acute{\psi}_{QRE_1}} \bra{\acute{\psi}_{QRE_1}} \right)\right] \otimes
\rho_R \otimes \rho_{E_1} \Big\} \\
& \leq & C_{E,f}({\mathcal E}_2).
\enqa
Hence,
\beq
C_{E,f}({\mathcal E}_2 \circ {\mathcal E}_1) \leq
\min\left\{C_{E,f}({\mathcal E}_1), C_{E,f}({\mathcal E}_2)\right\},
\enq
which is the data processing inequality.

\subsection{Coherent Information}

Let $Q$ be our system of interest with density matrix $\rho$ that is purified in the composite
system $RQ$. We consider two CPTP quantum operations ${\mathcal{E}_1}(\cdot)$ and
${\mathcal{E}_2}(\cdot)$ that are mocked up by introducing ancillae $E_1$ and $E_2$
respectively. The system after the operation ${\mathcal{E}_1}(\cdot)$ is denoted by
$R^{'} Q^{'} E_1^{'} E_2{'}$ and after the operation ${\mathcal{E}_2}(\cdot)$ by
$R^{''} Q^{''} E_1^{''} E_2{''}$.

The $f$-generalization of the coherent information is defined as
\beqa
I_f(\rho,{\mathcal E}_1) = -S_f\left( \rho_{R^{'} E_1^{'}} \Big|\Big|
{\mathrm{I}} \otimes \rho_{E_1^{'}} \right),
\enqa
where $\rho_{R^{'} E_1^{'}} = \tr_{Q^{'} E_2^{'}}
(\rho_{R^{'} Q^{'} E_1^{'} E_2{'}})$ and
$\rho_{E_1^{'}} = \tr_{R^{'} Q^{'} E_2^{'}}
(\rho_{R^{'} Q^{'} E_1^{'} E_2{'}})$. For $f(t) = -\ln(t)$, it was
shown by Shor that coherent information is related to the quantum channel capacity
\cite{shor-cap-2002}.

\begin{lemma}
For CPTP quantum operations ${\mathcal E}_1(\cdot)$ and ${\mathcal E}_2(\cdot)$, and
strictly positive density matrix $\rho$,
\beq
\label{dpinq3}
S_f(\rho) \geq I_f(\rho,{\mathcal E}_1) \geq I_f(\rho,{\mathcal E}_2 \circ {\mathcal E}_1).
\enq
For a non-affine and diffused operator convex function $f(\cdot)$, the equality holds
in the first inequality if and only if there exists a CPTP quantum operation
${\mathcal E}_2(\cdot)$ such that $F(\rho,{\mathcal E}_2 \circ {\mathcal E}_1) = 1$.
\end{lemma}
\begin{proof}
We first prove the inequalities. Since ${\mathcal{E}_1}(\cdot)$ does not affect $R$, hence,
$\rho_R = \rho_{R^{'}}$. Similar reasoning for ${\mathcal{E}_2}(\cdot)$ yields
$\rho_{R^{'} E_1^{'}} = \rho_{R^{''} E_1^{''}}$. We have using monotonicity
\beqa
S_f(\rho) & = & -S_f(\rho_R || {\mathrm{I}}) \\
& = & -S_f \left(\rho_{R^{'}} || {\mathrm{I}} \right) \\
\label{duminy1}
& \geq & -S_f \left(\rho_{R^{'}E_1^{'}} || {\mathrm{I}} \otimes \rho_{E_1^{'}}\right) \\
& = & I_f(\rho,{\mathcal E}_1)
\enqa
and
\beqa
I_f(\rho,{\mathcal E}_1) & = &
-S_f \left(\rho_{R^{'}E_1^{'}} || {\mathrm{I}} \otimes \rho_{E_1^{'}}\right) \\
& = & -S_f \left(\rho_{R^{''}E_1^{''}} ||
{\mathrm{I}} \otimes \rho_{E_1^{''}} \right) \\
\label{duminy2}
& \geq & -S_f \left(\rho_{R^{''}E_1^{''}E_2^{''}} ||
{\mathrm{I}} \otimes \rho_{E_1^{''}E_2^{''}} \right) \\
& = & I_f(\rho,{\mathcal E}_2 \circ {\mathcal E}_1),
\enqa
where Eqs. (\ref{duminy1}) and (\ref{duminy2}) follow using Eq. (\ref{equality1}).

To prove the equality condition, let us first assume that
$F(\rho,{\mathcal E}_2 \circ {\mathcal E}_1) = 1$. This implies that
\beq
\bra{\psi_{RQ}} \rho_{R^{''}Q^{''}} \ket{\psi_{RQ}} = 1,
\enq
and since $\rho_{R^{''}Q^{''}}$ is a density matrix, hence, it follows that
$\bra{\psi_{RQ}} \rho_{R^{''}Q^{''}} \ket{\psi_{RQ}} \leq 1$ with equality if
and only if $\ket{\psi_{RQ}}$ is an eigenvector of $\rho_{R^{''}Q^{''}}$
with eigenvalue $1$, or
\beq
\rho_{R^{''}Q^{''}} = \ket{\psi_{RQ}} \bra{\psi_{RQ}}.
\enq
Since $\rho_{R^{''}Q^{''}}$ is in a pure state, hence, it follows that
\beq
\rho_{R^{''}Q^{''}E_1^{''}E_2^{''}} = \ket{\psi_{RQ}} \bra{\psi_{RQ}} \otimes
\rho_{E_1^{''}E_2^{''}}
\enq
and by tracing out $Q^{''}$, we get
\beq
\rho_{R^{''}E_1^{''}E_2^{''}} = \rho_R \otimes \rho_{E_1^{''}E_2^{''}}.
\enq
Hence,
\beqa
I_f(\rho,{\mathcal E}_2 \circ {\mathcal E}_1) & = & -S_f
\left(\rho_{R^{''}E_1^{''}E_2^{''}} || {\mathrm{I}} \otimes \rho_{E_1^{''}E_2^{''}}\right) \\
& = & -S_f\left(\rho_R \otimes \rho_{E_1^{''}E_2^{''}}
|| {\mathrm{I}} \otimes \rho_{E_1^{''}E_2^{''}}\right) \\
& = & -S_f(\rho_R || {\mathrm{I}}) \\
\label{dummy7}
& = & S_f(\rho).
\enqa
Result follows by using Eq. (\ref{dpinq3}) and Eq. (\ref{dummy7}) to get
\beq
S_f(\rho) \geq I_f(\rho,{\mathcal E}_1) \geq I_f(\rho,{\mathcal E}_2 \circ {\mathcal E}_1)
= S_f(\rho).
\enq
To prove the statement in the other direction, let us assume that
\beq
I_f(\rho,{\mathcal E}_1) = S_f(\rho).
\enq
This implies that
\beq
S_f
\left(\rho_{R^{'}E_1^{'}} || {\mathrm{I}} \otimes \rho_{E_1^{'}}\right) 
= S_f \left(\rho_{R^{'}} || {\mathrm{I}} \right).
\enq
Using Eq. (\ref{equality1}), it follows that the above equality is true if and only if
\beq
\rho_{R^{'}E_1^{'}} = \rho_{R^{'}} \otimes \rho_{E_1^{'}},
\enq
which is the same condition as the one for $f(t) = -\ln(t)$ in
Ref. \cite{schumacher-nielsen-1996}, following which, we can construct a recovery
operation ${\mathcal{E}}_2(\cdot)$ such that
$\rho_{R^{''} Q^{''}} = \rho_{RQ}$ or $F(\rho,{\mathcal E}_2 \circ {\mathcal E}_1) = 1$.
\end{proof}

\section{Conclusions and Acknowledgements}

In conclusion, we have studied the fundamental properties of the \frel and the \fent. We give
the equality conditions under some properties for a class of operator convex functions. These
conditions are more general than the previously known conditions and also
apply to the case of $f(t) = -\ln(t)$. We define the \fent in terms of the \frel and study its
properties giving the equality conditions in some cases. We also show that the
$f$-generalizations of many well-known information-theoretic quantities also
satisfy the data processing inequality and for the case of $f$-coherent information,
we give the equality conditions.

The author thanks R. Bhatia for useful discussions on the operator convex functions.

\bibliographystyle{IEEEtran}
\bibliography{master}

\end{document}